\documentclass[%
 reprint,
superscriptaddress,
 amsmath,amssymb,
 aps,
prb,
]{revtex4-2}

\usepackage{graphicx}
\usepackage{dcolumn}
\usepackage{bm}

\usepackage{graphicx,epsf,bm,bbm,color,amssymb,float, subfigure}
\usepackage{amsmath,bm}
\usepackage{color}
\usepackage{textcomp}
\usepackage{dsfont}
\usepackage{tikz}
\usepackage{indentfirst}
\usepackage[colorlinks,citecolor=blue]{hyperref}
\usepackage{soul}
\usepackage{multirow}
\newcommand{\remove}[1]{}

\newcommand{\supplementarysection}{%
  \setcounter{figure}{0}
  \let\oldthefigure\thefigure
  \renewcommand{\thefigure}{S\oldthefigure}
  \setcounter{section}{0}
  \let\oldthesection\thesection
  \renewcommand{\thesection}{S\oldthesection}
  \setcounter{equation}{0}
  \let\oldtheequation\theequation
  \renewcommand{\theequation}{S\oldtheequation}
  \setcounter{table}{0}
  \let\oldthetable\thetable
  \renewcommand{\thetable}{S\oldthetable}
}

\newcommand{\bi}{\begin{itemize}}
\newcommand{\ei}{\end{itemize}}
\newcommand{\be}{\begin{enumerate}}
\newcommand{\ee}{\end{enumerate}}

\newenvironment{dfn}{{\vspace*{1ex} \noindent \bf Definition }}{\vspace*{1ex}}

\newcommand{\nn}{\nonumber}  %

\newcommand{\Eq}[1]{Eq.~(\ref{#1})}

\newcommand{\ket}[1]{\left| #1 \right>} 
\newcommand{\trace}[1]{\mathrm{tr}{\left(#1\right)}}
	\newcommand{\bra}[1]{\left< #1 \right|} 
	
	\newcommand{\beq}{\begin{eqnarray}}
	\newcommand{\eeq}{\end{eqnarray}}
	\newcommand{\bea}{\begin{eqnarray}\begin{aligned}}
	\newcommand{\eea}{\end{aligned}\end{eqnarray}}

\definecolor{forestgreen}{RGB}{34,139,34}

\begin{document}

\title{Charge-$6e$ superconductivity from doping $SU(3)$ spin liquids}

\author{Yan-Qi Wang}
\affiliation{Department of Physics and Joint Quantum Institute, University of Maryland, College Park, Maryland 20742, USA}
\author{Boran Zhou}
\affiliation{Department of Physics and Astronomy, Johns Hopkins University, Baltimore, Maryland 21218, USA}
\author{Hui Yang}
\email{huiyang.physics@gmail.com}
\affiliation{Department of Physics and Astronomy, University of Pittsburgh, Pennsylvania 15213, USA}
\author{Zhi-Qiang Gao}
\email{zqgao@berkeley.edu}
\affiliation{Department of Physics, University of California, Berkeley, California 94720, USA}

\begin{abstract}
We propose doping $SU(3)$-symmetric spin liquids as a route toward charge-$6e$ superconductivity. This generalizes the idea of constructing charge-$4e$ superconductivity from doped $SU(4)$-symmetric phases. As a concrete platform, we study a bilayer triangular-lattice Hubbard model with $SU(3)$ spin symmetry and interlayer antiferromagnetic exchange. Using complementary parton constructions, we analyze doped $\mathbb{Z}_3$ quantum spin liquid and $SU(3)$-related chiral spin liquids. Doping a $\mathbb{Z}_3$ quantum spin liquid can produce an orthogonal metal with a gauge invariant fermi surface of charge-$3e$ fermionic trions. Pairing these trions gives a time-reversal-symmetric charge-$6e$ superconductor. Doping Abelian $SU(3)_1$ and $SU(6)_1$ chiral spin liquids yields chiral charge-$6e$ superconductors with and without residual Abelian topological order, respectively. Doping a non-Abelian $SU(3)_2$ chiral spin liquid leads to a non-Abelian chiral charge-$6e$ superconductor intertwined with $SO(3)_{-3}$ topological order and supporting non-Abelian $h/(6e)$ superconducting vortices. We also identify several other phases, including $\mathbb{Z}_3$ orthogonal metal, quantum anomalous Hall (crystal) phases enriched by $\mathbb{Z}_3$ or $\mathbb{Z}_2$ topological order, $SU(3)$-breaking charge-$2e$ superconductors, composite fermi liquid coupled to non-Abelian gauge field, and descendant chiral spin liquids. Our results identify doped $SU(3)$ spin liquids as a natural setting where symmetry, fractionalization, and topology cooperate to produce charge-$6e$ superconductivity.
\end{abstract}

\maketitle

\section{Introduction}

Superconductivity (SC) is conventionally associated with the condensation of charge-$2e$ Cooper pairs.  More generally, however, SC only requires the spontaneous breaking of the global electromagnetic $U(1)$ symmetry, and the lowest-charge local operator with off-diagonal long-range order need not be a Cooper pair. A prominent example is charge-$4e$ SC, in which the elementary condensate is a quartet of electrons, while the charge-$2e$ Cooper pair does not acquire long-range order. Higher-charge SC has attracted considerable recent interest from several complementary viewpoints~\cite{Babaev2024,Zeng2024,Volovik2024, Soldini2024,Dai2024,Wu2024,samoilenka2025,zou2025,Shi2026, shi2025nonabelian,Gao2026,Zhang20254e,Gao20264e,Li20264e,Wan2026, Pan2026}. 

One widely studied route to higher-charge SC is through vestigial order of pair density wave states~\cite{Berg2009,Radzihovsky2009, Radzihovsky2011} or multicomponent SC~\cite{Vidal2002,Babaev2004,Radzihovsky2004,Radzihovsky2008, Herland2010,Zhou2021,Jian2021,Fernandes2021,Grinenko2021,Gao2022,Babaev2024,Zeng2024,Volovik2024,Soldini2024,Dai2024,Wu2024,samoilenka2025,zou2025,Shi2026}. In these settings, the higher-charge SC order parameter is often a composite of more conventional charge-$2e$ Cooper pairs. For example, fluctuations may disorder the primary charge-$2e$ pair field while preserving a composite charge-$4e$ order parameter. Related mechanisms arise in multicomponent SC, where the relative phases between several condensates can fluctuate strongly enough that only a higher-charge composite order remains coherent. These examples show that higher-charge SC is not forbidden by general principles, but many of them start from an already superconducting or nearly superconducting parent state.

A complementary route is provided by topology and fractionalization. In anyon superconductivity~\cite{FetterHannaLaughlin,Laughlin1988,Lee1989, WilczekWittenHalperinAnyonSC,Divic2025,Pichler2025,Senthil2025,
Shi2025,Kuhlenkamp2025,shi2025nonabelian}, the mobile dopants are fractionally charged anyons rather than electrons. The lowest-charge local boson that can acquire phase coherence is then determined not only by electric charge, but also by the fractional statistics and symmetry quantum numbers of the doped anyons. Consequently, the resulting SC order can have charge larger than $2e$. In simple cases, if the doped anyon has electric charge $2ne/q$ and $q$ such anyons fuse to a local boson, the resulting SC phase has charge $2ne$. This mechanism also naturally allows SC to coexist with deconfined anyons, leading to exotic higher-charge SC with Abelian or non-Abelian topological order and, in the non-Abelian case, non-Abelian SC vortices~\cite{shi2025nonabelian,Gao2026}. 

Symmetry provides another powerful organizing principle. In studies of charge-$4e$ SC, $SU(4)$ symmetry has long been invoked as a mechanism that favors electron quartets and suppresses ordinary Cooper pair condensation~\cite{Wu2005,Lecheminant2005,Capponi2007,Capponi2008,Roux2008,Ko2008,Zhang2020,Khalaf2022,Zhang20254e,Gao20264e,Li20264e,Wan2026,Pan2026}. More generally, it is recently proposed that, the nontrivial group center $\mathbb{Z}_N$ of $SU(N)$ symmetry can enforce higher charge singlet SC~\cite{Gao20264e}, and $SU(4)$ exemplifies a simplest application of such a center enforcement mechanism. To be concrete, the $\mathbb{Z}_N$ center enforces the lowest-charge singlet bound state of electrons to be composed by $N$ electrons. Consequently, any lower-charge condensate must spontaneously break the $SU(N)$ symmetry, which is forbidden at finite temperature according to Mermin--Wagner theorem. The $SU(4)$ case gives the simplest even-$N$ realization of this center enforcement mechanism: the minimal charged $SU(4)$-singlet bosonic operator is naturally a charge-$4e$ quartet, which can boost high-temperature charge-$4e$ SC in $SU(4)$-symmetric systems~\cite{Wan2026,Li20264e}. The $SU(3)$ case, which is the focus of this work, is more subtle and leads naturally to charge-$6e$ SC. Three electrons can form an $SU(3)$ singlet trion, but this charge-$3e$ trion is fermionic and hence cannot itself serve as an SC order parameter. The minimally charged bosonic singlet is instead a pair of trions, carrying charge $6e$. Thus, although having higher charge, a charge-$6e$ SC can be favored by a smaller symmetry than charge-$4e$ one, implying that it to be easier realized. This provides a symmetry-based motivation for studying doped $SU(3)$-symmetric states as a route to charge-$6e$ SC. Relatedly, emergent $U(6)\supset SU(3)$ symmetry has been proposed in $M$-point twisting Moir\'e materials~\cite{Clugru2025,Vasiliou2026,Clugru2026}, and in systems with \emph{threefold} rotational symmetry, charge-$6e$ SC has been reported to be more robust than charge-$4e$ SC~\cite{Zhou2021,Ge2024}.

Spin liquids provide a natural setting in which the symmetry-based and topological mechanisms discussed above are intertwined.  In $SU(N)$-symmetric Mott systems, two important classes of symmetric spin liquids are particularly relevant for higher-charge SC: the $\mathbb{Z}_N$ quantum spin liquid (QSL) and the chiral spin liquids (CSL). The former are described by discrete $\mathbb{Z}_N$ gauge fields and realize $\mathbb{Z}_N$ topological order. The latter spontaneously break time-reversal symmetry and realize lattice analogues of fractional quantum Hall states, whose universal topological responses are captured by Chern--Simons (CS) theories of emergent gauge fields. Doping spin liquids has emerged as a promising route toward unconventional SC~\cite{Divic2025,Pichler2025}. In particular, doping $\mathbb{Z}_4$ QSL~\cite{Zhang2020} and $SU(4)_1$ CSL~\cite{Zhang20254e} in $SU(4)$-symmetric systems can lead to charge-$4e$ SC, possibly coexisting with Abelian topological order.  

In this work, we generalize this paradigm to $SU(3)$-symmetric systems.  The $\mathbb{Z}_3$ center structure of $SU(3)$ changes the minimal symmetry-preserving charged boson from an electron quartet to a pair of trions, and therefore naturally points to charge-$6e$ rather than charge-$4e$ SC.  We study this idea in a bilayer $SU(3)$ Hubbard model, which provides a microscopic platform for both $\mathbb{Z}_3$ QSLs and $SU(3)$-related CSLs.  We show that doping the $\mathbb{Z}_3$ QSL leads to a ``conventional'' route toward charge-$6e$ SC by pairing the charge-$3e$ trions. We further show that doping Abelian $SU(3)_1$ and $SU(6)_1$ CSLs leads to chiral charge-$6e$ SC, with or without intrinsic Abelian topological order. Although doping an $SU(3)_1$ CSL has been considered previously~\cite{Zhang2026}, the emergence of charge-$6e$ SC was not the central focus. We also construct a non-Abelian version by doping an $SU(3)_2$ CSL.  In this case, the resulting phase is a charge-$6e$ SC intertwined with non-Abelian topological order and non-Abelian SC vortices. Thus, doped $SU(3)$ spin liquids provide a unified framework for realizing charge-$6e$ SC from both $\mathbb{Z}_3$ gauge structure and chiral topological order.

The paper is organized as follows.  In Sec.~\ref{sec:model} we introduce the bilayer Hubbard model with $SU(3)$ symmetry as a microscopic platform for $\mathbb{Z}_3$ QSL and $SU(3)$-related CSL.  In Sec.~\ref{sec:z3} we discuss doping states that preserve the layer symmetry, including the $\mathbb{Z}_3$ QSL state.  In Sec.~\ref{sec:csl} we discuss doping states that preserve the $SU(3)$ spin symmetry, including Abelian $SU(3)_1$ and $SU(6)_1$ CSL states~\cite{Lai2013,Zhang2026}, as well as a non-Abelian $SU(3)_2$ CSL state.  In Sec.~\ref{sec:dis} we conclude with discussion and outlook.

\section{The model}\label{sec:model}

Bilayer Hubbard model with interlayer Heisenberg interaction has been constructed to describe charge-$2e$ SC in nickelates with $SU(2)$ symmetry~\cite{Oh2024,Yang2025,Oh2025,ESDreview}, and primary charge-$4e$ SC with $SU(4)$ symmetry~\cite{Gao20264e}. Here we generalize this series of models to $SU(3)$ case. Consider the following $SU(3)$ symmetric model on the triangular lattice:
\beq
H&=&-t\sum_{\left<\mathbf{i}\mathbf{j}\right>}c_{l,\sigma}^\dagger(\mathbf{i})c_{l,\sigma}(\mathbf{j})+\frac{u}{2}\sum_{\mathbf{i},l}n^2_{l}(\mathbf{i})\nn\\
& &+\sum_{\mathbf{i}}v\,n_+({\bf i})n_-({\bf i})+J_\perp T^m_+(\mathbf{i})T^m_-(\mathbf{i}).\label{eq:bH}
\eeq
Here, $c_{l,\sigma}^\dagger(\mathbf{i})$ creates an electron in layer $l = \pm$ along with $SU(3)$ spin index $\sigma = 1,2,3$, where in each layer they are transformed under fundamental spinor $\mathbf{3}$ irrep of $SU(3)$. The positive $t$ and $u$ describes intralayer hopping and Hubbard interaction, respectively. The interlayer interactions consist of the density-density interaction described by $v$, and the $SU(3)$ antiferromagnetic Heisenberg coupling associated to positive $J_\perp$. $SU(3)$ generators in each layer are defined as $T_\pm^m(\mathbf{i})=(1/2)c^\dagger_{\pm,\sigma}(\mathbf{i})\lambda^m_{\sigma\sigma^\prime}c_{\pm,\sigma^\prime}(\mathbf{i})$, where $\lambda^{m=1,2,\cdots,8}$ are Gell-Mann matrices. The bilayer model \Eq{eq:bH} has $(SU(3)\times U(1)_+\times U(1)_-)/\mathbb{Z}_3$ symmetry. 

In the Mott limit, $u\gg J_\perp,v\gg t$. The low-energy Hilbert space of model \Eq{eq:bH} is spanned by local states on each rung connecting the two layers. In what follows, we use $x$ to measure the doping level away from half-filling, where $x=0$ and $x=1$ are corresponding to three and two electrons per rung, respectively. At half filling $x=0$, according to the antiferromagnetic Heisenberg interaction, the lowest energy local states are two $SU(3)$ singlet trion states $\ket{f_\pm (\mathbf{i})}=(1/\sqrt{6})\epsilon_{\mu\nu\rho}c^\dagger_{\pm,\mu}(\mathbf{i})c^\dagger_{\pm,\nu}(\mathbf{i})c^\dagger_{\mp,\rho}(\mathbf{i})\ket{0(\mathbf{i})}$. These two states are degenerate, suggesting an emergent $SU(2)_l$ layer symmetry at low energy, where $(\ket{f_+ (\mathbf{i})},\ket{f_- (\mathbf{i})})$ form its fundamental spinor. The effective Hamiltonian is,
\beq
H_{x=0}=J_\mathrm{eff}\sum_{\left<\mathbf{ij}\right>}\mathbf{S}(\mathbf{i})\cdot\mathbf{S}(\mathbf{j})-\frac{1}{4}n_f(\mathbf{i})n_f(\mathbf{j})+\cdots,\label{eq:x0}
\eeq
where $\mathbf{S}(\mathbf{i})=(1/2)\ket{f_l (\mathbf{i})}{\bm \tau}_{ll^\prime}\bra{f_{l^\prime}(\mathbf{i})}$ is the $SU(2)_l$ generators defined by Pauli matrices $\tau^{1,2,3}$. The effective Hamiltonian \Eq{eq:x0} indeed has $SU(2)_l$ symmetry up to the order of $J_\mathrm{eff}\sim t^2/J_\perp$, while higher order terms break it down to $U(1)_z$. 

At $1/3$ filling $x=1$, there are two electrons per rung, where the Hubbard interaction further enforces one electron per site. The lowest energy local states form $\bar{\mathbf{3}}$ under $SU(3)$ and singlet under $SU(2)_l$, $\ket{h_\sigma (\mathbf{i})}=(1/\sqrt{2})\epsilon_{\mu\nu\rho}c^\dagger_{+,\nu}(\mathbf{i})c^\dagger_{-,\rho}(\mathbf{i})\ket{0(\mathbf{i})}$. The three $\ket{h_\sigma (\mathbf{i})}$ states form the fundamental anti-spinor $\bar{\mathbf{3}}$ irrep of $SU(3)$ arising from decomposition $\mathbf{3}\otimes\mathbf{3}=\mathbf{6}\oplus\bar{\mathbf{3}}$. The effective Hamiltonian at $x=1$ has leading term at the order of $t^2/u$ composed by bond and ring exchanges~\cite{Lai2013} defined in $SU(3)$ generators $\mathbf{T}(\mathbf{i})=-(1/2)\ket{h_\sigma (\mathbf{i})}{\bm \lambda}_{\sigma\sigma^\prime}^*\bra{h_{\sigma^\prime}(\mathbf{i})}$. 

For generic doping $0\le x\le 1$, we keep five states in the low energy local Hilbert space, the $SU(3)$ singlet and $SU(2)_l$ spinor $\ket{f_\pm (\mathbf{i})}$, and the $SU(2)_l$ singlet $SU(3)$ anti-spinor $\ket{h_\sigma (\mathbf{i})}$. The system has an emergent $SU(2)_l\times SU(3)$ symmetry at low energy.

\section{Doping layer symmetric states}\label{sec:z3}

In this section we consider doping states preserving layer $SU(2)_l$ symmetry. Here the electron is naturally decomposed into partons $f_l(\mathbf{i})$ and $h_\sigma(\mathbf{i})$ corresponding to the states in the local Hilbert space,
\beq
c_{l,\sigma}(\mathbf{i})=h_\sigma^\dagger(\mathbf{i})f_l(\mathbf{i}),\label{eq:parA}
\eeq
with densities $n_f(\mathbf{i})=1-x$ and $n_h(\mathbf{i})=x$. As $f_l$ is fermionic, it does not condense, which in general preserves the layer symmetry. The $SU(2)_l$ and $SU(3)$ generators read,
\beq
\mathbf{S}(\mathbf{i})=\frac{1}{2}f^\dagger_l(\mathbf{i}){\bm \tau}_{ll^\prime}f_{l^\prime}(\mathbf{i}),\quad \mathbf{T}(\mathbf{i})=-\frac{1}{2}h^\dagger_\sigma(\mathbf{i}){\bm \lambda}_{\sigma\sigma^\prime}^* h_{\sigma^\prime}(\mathbf{i}).
\eeq
A generic Lagrangian describing low energy physics is,
\beq
\mathcal{L}[c;A]=\mathcal{L}[f;a+A]+\mathcal{L}[h;a],\label{eq:L2}
\eeq
where $a$ is a dynamical $U(1)$ gauge field characterizing the gauge redundancy in the parton construction \Eq{eq:parA}, and $A$ is the background electromagnetic field. As it couples to fermion $f_l$, it is a spin$_\mathbb{C}$ connection, instead of an ordinary $U(1)$.

There are three main possibilities of state formed by $f_l$ to preserve $SU(2)_s$ symmetry: fermi surface, $SU(2)_1$ chiral layer liquid, or $\mathbb{Z}_2$ layer liquid where $f_l$ form layer-singlet Cooper pairs~\cite{DopeRMP}. Upon doping, the resulting phase depends on the state formed by the holon $h_\sigma$.

\subsection{Doped holons are gapped}

When the doped holons are gapped, it effectively decouples from the low energy physics. In this case, all the continuous and the lattice symmetries are preserved. When $f_l$ forms a $\mathbb{Z}_2$ layer liquid or $SU(2)_1$ chiral layer liquid state, the system remains in such a phase. When $f_l$ exhibits fermi surface, the system realizes an orthogonal metal phase~\cite{Nandkishore2012}.

\subsection{Doped holons are condensed}

Condensation of doped holons is likely to happen at large doping $x\approx 1$, where the $SU(3)$ symmetry is spontaneously broken to $SU(2)$, and the dynamical $U(1)$ gauge field $a$ is Higgsed. It also identifies $c_{l,\sigma}\sim f_l$, and low energy effective theory $\mathcal{L}[A]=\mathcal{L}[f;A]$. Therefore, the fermi surface, the $SU(2)_1$ chiral layer liquid states of $f_l$, and the $\mathbb{Z}_2$ layer liquid with singlet-paired $f_l$ translates to the fermi liquid, the $C=2$ quantum anomalous Hall insulator, and the charge-$2e$ SC phases of electron. These phases are all coupled to gapless Goldstone modes arising from $SU(3)$ breaking. Therefore, they only survive at zero temperature, according to the Mermin--Wagner theorem. In particular, in the fermi liquid phase, the coupling between fermi surface and critical boson may drive a non-fermi liquid phase, such as pseudogap metal or strange metal, at finite temperature with disorder~\cite{DopeRMP,Metlitski2010,Sachdev2010,Hartnoll2011,Patel2014,Sachdev2016,ZhangSachdev,Patel2023}.

\subsection{Doped holons form $\mathbb{Z}_3$ quantum spin liquid}

A more interesting scenario is that the condensed object is not a single holon $h_\sigma$, but a bound state of it. In particular, the $\mathbb{Z}_3$ center of $SU(3)$ enforces a pair of $h_\sigma$ to remain charged under $SU(3)$, while three $h_\sigma$ can form a singlet trimer $\epsilon_{\mu\nu\rho}h_\mu(\mathbf{i})h_\nu(\mathbf{j})h_\rho(\mathbf{k})$. Condensation of such a trimer gives rise to a $SU(3)$ symmetric $\mathbb{Z}_3$ quantum spin liquid state. In principle, such a condensation can be favored at intermediate doping $0<x<1$ by an $SU(3)$ ring exchange on triangular plaquettes that does not explicitly break time reversal symmetry~\cite{Lai2013}. A similar fully symmetric $\mathbb{Z}_4$ quantum spin liquid state has been proposed in $SU(4)$ Hubbard model~\cite{Zhang2020}. In the $\mathbb{Z}_3$ quantum spin liquid state, the low energy effective theory reads,
\beq
\mathcal{L}[c;A]=\mathcal{L}[f;a+A]+\frac{3}{2\pi}\tilde{a}\mathrm{d}a,
\eeq
where $\tilde{a}$ is the hydrodynamic field Higgsing $a$ to a $\mathbb{Z}_3$ gauge field. When $f_l$ forms a fermi surface, the condensed holon trimer glues the physical electron into a charge-$3e$ trion, $F\sim \left<\epsilon_{\mu\nu\rho}h_\mu h_\nu h_\rho\right>f_l f_{l^\prime} f_{l^{\prime\prime}}$, which is a gauge invariant gapless excitation~\cite{Xu2023,Stepp2026}. The charged fermi surface coupled to dynamical $\mathbb{Z}_3$ gauge field realizes an orthogonal metal phase~\cite{Nandkishore2012} of trion. Its fermi surface volume is one third of that in a non-interacting fermi liquid, which suggests its strong correlation nature. It is similar to the $\mathbb{Z}_3$ orthogonal metal~\cite{Shi2026sc,Senthil2026,Han20261,Zhang20261,Mehta2026} proposed to explain the metallic phase observed in twisted MoTe$_2$~\cite{Xu2025}; however, that $\mathbb{Z}_3$ orthogonal metal phase has gapless excitation with fractional electric charge $e/3$, which is different from ours with integer charge $e$.

When $f_l$ forms $SU(2)_1$ chiral layer liquid, the topological response is,
\beq
\mathcal{L}[A]=\frac{2}{4\pi}(a+A)\mathrm{d}(a+A)+\frac{3}{2\pi}\tilde{a}\mathrm{d}a+4\Omega_g,\label{eq:LZ3}
\eeq
where $\Omega_g$ is the gravitational CS term. It appears because $(a+A)$ is a dynamical spin$_\mathbb{C}$ gauge field. To be precise, a level-1 CS term of a dynamical spin$_\mathbb{C}$ gauge field is accompanied with $2\Omega_g$, and each $\Omega_g$ contributes chiral central charge $1/2$. Lagrangian \Eq{eq:LZ3} describes a $\mathbb{Z}_3$ topological order subject to a Dijkgraaf--Witten twist~\cite{DWT}. The vison in this phase whose current couples to $\tilde{a}$ has electric charge $2e/3$ and topological spin $1/9$. However, this twisted $\mathbb{Z}_3$ topological order does not contribute to the Hall conductance $\sigma_H=2$ and chiral central charge $c_-=2$ of this quantum anomalous Hall (QAH) phase. Thus, this phase can be viewed as an QAH$^*$.

When $f_l$ forms a $\mathbb{Z}_2$ quantum layer liquid with its Cooper pair condensed, $\left<\epsilon_{ll^\prime}f_l(\mathbf{i})f_{l^\prime}(\mathbf{j})\right>\neq 0$, the dynamical spin$_\mathbb{C}$ gauge field $(a+A)$ is Higgsed to $\mathbb{Z}_2$. However, as $a$ is already Higgsed to $\mathbb{Z}_3$ by the holon trimer, the $U(1)$ charge conservation should actually be broken to $\mathbb{Z}_{6=\mathrm{lcm}(2,3)}$, suggesting a charge-$6e$ SC. To see this, consider the topological response,
\beq
\mathcal{L}[A]&=&\frac{2}{2\pi}b\mathrm{d}(a+A)+\frac{3}{2\pi}\tilde{a}\mathrm{d}a.
\eeq
Redefine $(\alpha,\tilde{\alpha})=(2b+3\tilde{a},-b-\tilde{a})$ as a legal $GL(2,\mathbb{Z})$ transformation and integrating out $\alpha$ yields a topological response
\beq
\mathcal{L}[A]=\frac{6}{2\pi}\tilde{\alpha}\mathrm{d}A,
\eeq
which indeed describes a charge-$6e$ SC. Such a charge-$6e$ SC has  real space order parameter 
\beq
\Delta_{6e}&\sim &\left<\epsilon_{ll^\prime}f_l(\mathbf{i})f_{l^\prime}(\mathbf{j})\right>\left<\epsilon_{ll^\prime}f_l(\mathbf{j})f_{l^\prime}(\mathbf{k})\right>\left<\epsilon_{ll^\prime}f_l(\mathbf{k})f_{l^\prime}(\mathbf{i})\right>\nn\\
&&\times\left<\epsilon_{\mu\nu\rho}h^\dagger_\mu(\mathbf{i})h^\dagger_\nu(\mathbf{j})h^\dagger_\rho(\mathbf{k})\right>^2.\label{eq:D6e}
\eeq
It is a descendant state of the trion orthogonal metal, where the charge-$3e$ trions on the trion fermi surface form charge-$6e$ Cooper pairs. As it is spin-layer symmetric, this charge-$6e$ trion SC can persist to finite temperature in the BKT sense. Similarly, the spin-layer symmetric orthogonal metal and QAH$^*$ phases are also possible at finite temperature.

The charge-$6e$ SC descendant from the trion orthogonal metal is similar to the charge-$2e$ SC descendant from the $\mathbb{Z}_3$ orthogonal metal~\cite{Shi2026sc,Senthil2026,Han20261,Zhang20261,Mehta2026}. In the $\mathbb{Z}_3$ orthogonal metal, the elementary gapless excitation has charge $e/3$, whose pairing leads to an SC phase with charge $3\times 2\times e/3=2e$, similar to \Eq{eq:D6e}. In the trion orthogonal metal, however, the elementary gapless excitation has charge $e$, giving rise to a charge-$6e$ SC.

\section{Doping spin symmetric states}\label{sec:csl}

In this section we consider doping spin symmetric states. Such states are more conveniently described using an alternative parton construction,
\beq
c_{l,\sigma}(\mathbf{i})=z_l^\dagger(\mathbf{i})d_\sigma(\mathbf{i}),\label{eq:parB}
\eeq
where $d_\sigma(\mathbf{i})$ is an $SU(2)_l$ singlet fermionic parton transformed under $\mathbf{3}$ of $SU(3)$, and $z_l(\mathbf{i})$ is an $SU(3)$ singlet bosonic spinor of layer $SU(2)_l$. These partons are not necessarily related to the $f_l$ and $h_\sigma$ partons introduced before. The fermionic nature of $d_\sigma$ generally preserves the $SU(3)$ symmetry. The densities of the partons satisfy $n_z(\mathbf{i})=1-x$ and $n_d(\mathbf{i})=x$. Under \Eq{eq:parB}, the $SU(3)$ and the $SU(2)_l$ generators are written as,
\beq
\mathbf{S}(\mathbf{i})=\frac{1}{2}z^\dagger_l(\mathbf{i}){\bm \tau}_{ll^\prime}z_{l^\prime}(\mathbf{i}),\quad \mathbf{T}(\mathbf{i})=-\frac{1}{2}d_\sigma(\mathbf{i}){\bm \lambda}_{\sigma\sigma^\prime}^* d^\dagger_{\sigma^\prime}(\mathbf{i}).
\eeq
The low energy physics is described by the Lagrangian,
\beq
\mathcal{L}[c;A]=\mathcal{L}[z;a]+\mathcal{L}[d;a+A].\label{eq:L2}
\eeq
Here we assign physical electric charge to $d_\sigma$ to avoid dynamical spin$_\mathbb{C}$ gauge fields.

An important class of spin symmetric states of the bilayer Hubbard model \Eq{eq:bH} is the CSL states. At $x=1$, \Eq{eq:bH} realizes an $SU(3)$ spin model with Heisenberg and ring exchanges. It is proposed that, at $x=1$, a strong enough ring exchange favors CSL states~\cite{Lai2013}, and the type of CSL depends on the relative strength between the two exchange interactions. Thus, we will mainly study doping these CSL states formed by $d_\sigma$ at $x=1$. The $SU(3)_1$ or the $SU(6)_1$ CSL state is found when the emergent gauge flux on each plaquette is $\Phi=\pi/3$ or $\Phi=2\pi/3$, respectively~\cite{Lai2013,Boos2020}, and we verify both using parton mean-field calculation (see Appendix~\ref{app:B}). Doping the two CSL states are discussed in Sec.~\ref{sec:SU31} and Sec.~\ref{sec:SU61}. For $\Phi=2\pi/3$, another possibility is the non-Abelian $SU(3)_2$ Fibonacci CSL state~\cite{Timmel2023}, whose doping is discussed in Sec.~\ref{sec:SU32}.

Before proceeding, we briefly comment other possible $SU(3)$ symmetric states. Fermi surface state of $d_\sigma$ (see Appendix~\ref{app:B}) can realize more conventional layer-polarized or inter-layer-coherent phases when $z_l$ is further condensed. However, the pairing of $d_\sigma$ cannot yield an $SU(3)$ symmetric state, as the $\mathbb{Z}_3$ center of $SU(3)$ group forbids a singlet pairing.

\subsection{Doping $SU(3)_1$ chiral spin liquid}\label{sec:SU31}

The mean-field ansatz of the $SU(3)_1$ CSL is $(\nabla\times {\bm a})/(2\pi)=1/3$ and $n_{d,\sigma}(\mathbf{i})=1/3$, where each magnetic unit cell contains three sites, giving rise to three Hofstadter bands with $C=1$. Correspondingly, each $d_\sigma$ fermion fills the lowest band and form a Chern insulator. The effective theory is,
\beq
\mathcal{L}[c;A]=\mathcal{L}[z;a]+\sum_{\sigma}\frac{1}{2\pi}b_\sigma\mathrm{d}(a+A)-\frac{1}{4\pi}b_\sigma\mathrm{d}b_\sigma,\label{eq:CSLp}
\eeq
where $b_\sigma$ are dynamical $U(1)$ gauge fields describing the Chern insulators formed by $d_\sigma$. Without doping, integrating out $a$ and then $b_1$ gives rise to the standard form of an $SU(3)_1$ CSL as a Halperin-221 state:
\beq
\mathcal{L}[A]=-\frac{2}{4\pi}b_2\mathrm{d}b_2-\frac{2}{4\pi}b_3\mathrm{d}b_3-\frac{1}{2\pi}b_2\mathrm{d}b_3,
\eeq
which is charge-neutral. On the other hand, integrating out $b_\sigma$ formally suggests a $U(1)_3$ CS theory; however, it leaves gravitational CS term $\Omega_g$ accompanied with dynamical spin$_\mathbb{C}$ gauge field $(a+A)$:
\beq
\mathcal{L}[A]=\frac{3}{4\pi}(a+A)\mathrm{d}(a+A)+6\Omega_g.
\eeq
With inclusion of $\Omega_g$, the above two descriptions of $SU(3)_1$ CSL have the same chiral central charge $c_-=2$.

Upon doping, we assume $d_\sigma$ remains in the $C=1$ states. The Streda formula then yields $(\nabla\times \delta{\bm a})/(2\pi)=\delta n_{d,\sigma}/C=-(1-x)/3$. The bosonic parton that determines the low energy physics has density $n_z(\mathbf{i})=1-x$. Magnetic translation symmetry tripling the unit cell generates three degenerate valleys $z_l^{I=1,2,3}$ in the band structure of $z_l$ in the magnetic Brillouin zone. Therefore, each $z_l^I$ has density $(1-x)/6$, feels the emergent flux $(\nabla\times \delta{\bm a})/(2\pi)=-(1-x)/3$, and hence is at effective filling $\nu_z=-1/2$.

Generally, there are three main possibilities of the state formed by $z_l^I$, where $z$ is condensed, pair condensed, or topologically ordered as a fractional quantum Hall state. We first consider $z_l^I$ condensed state, in which both the layer $SU(2)_l$ symmetry and the translation symmetry are spontaneously broken. Condensation of $z_l^I$ Higgses dynamical gauge field $a$, and effective field theory \Eq{eq:CSLp} describes a Chern insulator with $C=3$. In total, this phase is a layer-polarized and spin-symmetric anomalus Hall crystal (AHC) with Hall conductivity and chiral central charge $\sigma_H=c_-=3$. Although $SU(2)_l$ is spontaneously broken, this state should be able to persist to finite temperature, as layer $SU(2)_l$ symmetry is not an exact symmetry of the system.

Pair condensation of $z_l^I$ can fall in either layer-singlet or layer-triplet channel. The singlet channel preserves the $SU(2)_l$ symmetry, while the triplet channel breaks it. The translation symmetry is broken, regardless of the pairing channel. Both two channels Higgs the dynamical $U(1)$ gauge field $a$ to $\mathbb{Z}_2$. The topological response of $z_l^I$ pair condensed state is,
\beq
\mathcal{L}[A]=\frac{2}{2\pi}a\mathrm{d}\tilde{a}+\frac{3}{4\pi}(a+A)\mathrm{d}(a+A)+6\Omega_g,
\eeq
where $\tilde{a}$ is the hydrodynamic field that Higgses $a$ to $\mathbb{Z}_2$. It has the same Hall conductance as the $z_l^I$ condensed phase, $\sigma_H=3$, with a coexisting fermionic $\mathbb{Z}_2$ topological order~\cite{Gu2014}. In particular, the vison, {\it i.e.}, the vortex of the pair condensate, whose current is coupled to $\tilde{a}$ has physical electric charge $-e/2$ and topological spin $3/8$. In total, the system is a spin-layer symmetric, or a spin-symmetric and layer-polarized, AHC$^*$ with $\sigma_H=c_-=3$, for the singlet or the triplet channel, respectively.

A more interesting possibility is that $z_l^I$ forms a fractional quantum Hall state symmetric under layer symmetry. As each $z_l^I$ has effective filling $\nu_z=-1/2$, a natural choice is that it realizes a bosonic Laughlin state at half filling. However, it is worth noticing that this state does not preserve the full emergent $SU(2)_l$ symmetry, but only its $U(1)_z$ subgroup generated by $\tau^3$, {\it i.e.}, the UV symmetry of the lattice model. The topological response is,
\beq
\mathcal{L}[A]&=&\sum_{l,I}\frac{2}{4\pi}\beta_l^I\mathrm{d}\beta_l^I+\frac{1}{2\pi}\beta_l^I\mathrm{d}a\nn\\
&&+\frac{3}{4\pi}(a+A)\mathrm{d}(a+A)+6\Omega_g,\label{eq:Lraw}
\eeq
where $\beta_l^I$ is the dynamical $U(1)$ gauge field describing the bosonic Laughlin state. An $GL(7,\mathbb{Z})$ transformation yields,
\beq
\mathcal{L}[A]&=&\frac{3}{4\pi}\alpha_6\mathrm{d}\alpha_6+\sum_{J=1}^5\frac{2}{4\pi}\alpha_J\mathrm{d}\alpha_J-\frac{1}{2\pi}\alpha_J\mathrm{d}(\alpha_6+A)\nn\\
&&+\frac{6}{2\pi}\tilde{\alpha}\mathrm{d}A+6\Omega_g,
\eeq
where $(\tilde{\alpha},{\bm \alpha})=(a,{\bm \beta})W^\mathbf{T}$ for $W\in GL(7,\mathbb{Z})$ (see Appendix~\ref{app:A}). The level-6 mutual CS term between $\tilde{\alpha}$ and $A$ Higgses $A$ to $\mathbb{Z}_6$, which suggests a charge-$6e$ SC. It coexists with an Abelian topological order with torus ground state degeneracy $16$ and chiral central charge $c_-=-6$ determined by the $K$-matrix. Thus, this state is a spin-layer symmetric~\footnote{Here only the UV symmetry layer $U(1)_z$ is strictly preserved} chiral charge-$6e$ SC$^*$ with chiral central charge $c_-=-3$.

\subsection{Doping $SU(6)_1$ chiral spin liquid}\label{sec:SU61}

The mean-field ansatz of the $SU(6)_1$ CSL is $(\nabla\times {\bm a})/(2\pi)=2/3$ and $n_{d,\sigma}(\mathbf{i})=1/3$. Here the three Hofstadter bands in the magnetic unit cell have $C=2$, leading to the effective theory~\footnote{The standard form of the $SU(6)_1$ CSL without doping can be similarly derived by first integrating out $a$ in \Eq{eq:L2}.},
\beq
\mathcal{L}[c;A]=\mathcal{L}[z;a]+\frac{6}{4\pi}(a+A)\mathrm{d}(a+A)+12\Omega_g,
\eeq
Similar to the doping $SU(3)_1$ CSL case, upon doping the bosonic parton has density $n_z(\mathbf{i})=1-x$, and magnetic translation symmetry also generates three degenerate valleys $z_l^{I=1,2,3}$ in the band structure of $z_l$. According to Streda formula, here $(\nabla\times \delta{\bm a})/(2\pi)=\delta n_{d,\sigma}/C=-(1-x)/6$. Therefore, the effective filling of each $z_l^I$ is now $\nu_z=-1$.

When $z_l^I$ is condensed, a layer-polarized and spin-symmetric AHC state with Hall conductivity $\sigma_H=6$ is produced. When $z_l^I$ is pair condensed, the system realizes a spin-layer symmetric or a spin-symmetric and layer-polarized AHC$^*$, coexisting with a double semion topological order and $\sigma_H=6$, depending on the pairing channel of singlet or triplet, respectively.

When $z_l^I$ is not condensed, as $z_l^I$ has filling $-1$, it is natural to have $z_+^I$ and $z_-^I$ in valley $I$ form a layer-singlet bosonic IQH state at total filling $-2$. The topological response is thereby, 
\beq
\mathcal{L}[A]&=&\sum_{I}\frac{1}{2\pi}\beta_+^I\mathrm{d}\beta_-^I+\frac{1}{2\pi}(\beta_+^I+\beta_-^I)\mathrm{d}a\nn\\
&&+\frac{6}{4\pi}(a+A)\mathrm{d}(a+A)+12\Omega_g,
\eeq
where $\beta_l^I$ is the dynamical $U(1)$ gauge field describing the bosonic IQH state. Integrating out $\beta_l^I$ yields,
\beq
\mathcal{L}[A]=\frac{6}{2\pi}a\mathrm{d}A+\frac{6}{4\pi}A\mathrm{d}A+12\Omega_g,
\eeq
which suggests a chiral charge-$6e$ SC with chiral central charge $c_-=6$. Different to the doping $SU(3)_1$ CSL case, the spin-layer symmetric charge-$6e$ SC here does not coexist with topological order.

For $z_l^I$ at effective filling $\nu=-1$, an alternative possibility is that they form a composite fermi liquid (CFL), in analogue to that formed by electrons in the half-filled Landau level~\cite{HLR,Son2015}. In such a state, the composite fermion is formed by $z_l^I$ attached to one flux. Formally, the effective field theory can be written as
\beq
\mathcal{L}[A]&=&\sum_{l,I}\mathcal{L}_\mathrm{FS}[\psi_l^I;\gamma_l^I]-\frac{1}{4\pi}(\gamma_l^I-a)\mathrm{d}(\gamma_l^I-a)\nn\\
&&+\frac{6}{4\pi}(a+A)\mathrm{d}(a+A),
\eeq
where $\psi_l^I$ is a 6-flavor composite fermion with the same density as $z_l^I$. Thus $\nabla\times{\bm \gamma}_l^I=0$, and $\psi_l^I$ sees no net flux. It then forms a fermi surface described by $\mathcal{L}_\mathrm{FS}[\psi_l^I;\gamma_l^I]$ and coupled to dynamical spin$_\mathbb{C}$ gauge field $\gamma_l^I$~\footnote{The gravitational CS term is canceled as $\gamma_l^I$ is spin$_\mathbb{C}$}. Shifting $\gamma_l^I \mapsto \gamma_l^I+A$ and $a\mapsto a-A$ converts $\gamma_l^I$ to ordinary $U(1)$ gauge field and $a$ to spin$_\mathbb{C}$ gauge field. Further integrating out $a$ yields a 5-by-5 $K$-matrix of $\gamma$ describing an $SU(6)_1$ CSL~\cite{Lai2013}. Thus, the effective field theory can be viewed as a CFL of 6-flavor fermion, transformed under $SU(6)$ fundamental spinor irrep, coupled to an $SU(6)_1$ CS theory:
\beq
\mathcal{L}[A]=\mathcal{L}_\mathrm{FS}[\psi;\gamma-A\mathbb{I}_6]-\frac{1}{4\pi}\trace{\gamma\mathrm{d}\gamma+\frac{2}{3}\gamma^3}.
\eeq
Pairing of $\psi$ is enhanced by gauge fluctuations, which is either symmetric or asymmetric under $SU(6)$~\footnote{Direct product of two $SU(6)$ fundamental spinor $\mathbf{6}$ can be decomposed into symmetric $\mathbf{21}$ or asymmetric $\mathbf{15}$. Symmetric pairing transforms in $\mathbf{21}$, while asymmetric pairing transforms under $\mathbf{15}$.}. Symmetric pairing Higgses $SU(6)$ to $SO(6)$, yielding an $SO(6)_1$ CSL with chiral central charge $c_-=3$, while asymmetric pairing Higgses $SU(6)$ to $Sp(6)$, suggesting an $Sp(6)_1$ CSL with chiral central charge $c_-=21/5$. In particular, $Sp(6)_1$ is non-Abelian. Both of the CSL phases can break the translation symmetry through inter-valley pairing, or the layer symmetry through inter-layer pairing. If, under proper physical situation, the leading instability on the composite fermi surface is $SU(6)$ singlet hextetting, the condensation of such an hextet will preserve the $SU(6)_1$ sector and break the $U(1)$ charge conservation to $\mathbb{Z}_6$, leading to a charge-$6e$ SC$^*$ with coexisting $SU(6)_1$ topological order and chiral central charge $c_-=5$.


\subsection{Doping $SU(3)_2$ non-Abelian chiral spin liquid}\label{sec:SU32}

Parton construction \Eq{eq:parB} is not applicable to describe non-Abelian CSL state, as it only introduces an Abelian $U(1)$ gauge redundancy. According to the level--rank duality~\cite{Hsin2016}, $SU(3)_2$ is dual to $U(2)_{-3,-6}$, as the two theories have the same anyon fusion and braiding~\footnote{To be precise, at spin TQFT level, the exact duality is $SU(3)_2\boxtimes \mathrm{sVec}$ dual to $U(2)_{-3,-6}+12\Omega_g$, where $\mathrm{sVec}=\{1,f\}$ describes a trivial invertible fermionic phase with transparent fermion $f$.}. Therefore, we modify the parton construction \Eq{eq:parB} to accommodate a $U(2)$ gauge redundancy,
\beq
c_{l,\sigma}(\mathbf{i})=z_{l,s}^\dagger(\mathbf{i})d_{s,\sigma}(\mathbf{i}),
\eeq
where $s=1,2$ denotes an internal $U(2)$ gauge index. The effective field theory is thereby,
\beq
\mathcal{L}[c;A]=\mathcal{L}[z;a]+\mathcal{L}[d;a+A\mathbb{I}_2],
\eeq
where $a$ is the $U(2)$ gauge field. As the flavor of $d$ is doubled, each $d_{s,\sigma}$ now forms a $C=1$ Chern insulator. The doublet $(d_{1,\sigma},d_{2,\sigma})$ for a given $\sigma$ is then described by a $U(2)_{-1,-2}$ CS theory which is topologically equivalent to a $C=2$ Chern insulator. The effective theory of the $d_{s,\sigma}$ parton is the $U(2)_{-3,-6}$ CS theory,
\beq
\mathcal{L}[d;a+A\mathbb{I}_2]&=&\frac{3}{4\pi}\trace{a\mathrm{d}a+\frac{2}{3}a^3}+\frac{3}{2\pi}(\mathrm{tr}\,a)\mathrm{d}A\nn\\
&&+\frac{6}{4\pi}A\mathrm{d}A+12\Omega_g,
\eeq
which indeed describes an $SU(3)_2$ CSL by shifting $a\mapsto a-A\mathbb{I}_2$. In this subsection we only consider $z_{l,s}$ uncondensed case.

To see the effective filling of $z_{l,s}$ upon doping, we decompose $a=\tilde{a}\mathbb{I}_2+a^j\tau^j/2$, where $\tilde{a}$ and $a^j$ describe the $U(1)$ and the $SU(2)$ sectors, respectively. Only the $U(1)$ piece $\tilde{a}$ is coupled to $A$, while the $SU(2)$ sector is neutral. At doping level $x$, the density change of $d_{\sigma,s}$ is then $-(1-x)/6$ to respect the $U(2)$ gauge structure. Streda formula then yields $(\nabla\times \delta\tilde{\bm a})/(2\pi)=\delta n_{d,\sigma,s}/C=-(1-x)/6$. As magnetic translation symmetry generates three valleys in the band structure of $z_{l,s}$, each valley of $z_{l,s}$ has density $(1-x)/12$. Thus, the effective filling of $z_{l,s}^I$ is $\nu_z=-1/2$. The state formed by $z_{l,s}^I$ should preserve both the $SU(2)_l$ symmetry and the $U(2)$ gauge structure. In addition, upon taking $a=\tilde{a}\mathbb{I}_2$, it should reproduce the Hall response of a bosonic Laughlin state at $\nu=-1/2$ for each flavor of $z^I_{l,s}$. The simplest choice is that each valley forms a state described by $U(2)_{-1,2}$ bosonic CS theory~\footnote{A $U(2)_{k_1,k_2}$ theory is bosonic only when $2k_1+k_2\in 8\mathbb{Z}$. Therefore, $U(2)_{-1,2}$ is indeed bosonic.},
\beq
\mathcal{L}[Z^I;a]=\frac{1}{4\pi}\trace{a\mathrm{d}a+\frac{2}{3}a^3}-\frac{1}{4\pi}(\mathrm{tr}\,a)\mathrm{d}(\mathrm{tr}\,a),
\eeq
where $Z^I$ denotes the quartet of $z^I_{l,s}$ in valley $I$. The full $\mathcal{L}[z;a]$ is equal to $3\mathcal{L}[Z^I;a]$. Clearly, for $a=\tilde{a}\mathbb{I}_2$, $\mathcal{L}[Z^I;a]=-2/(4\pi)\,a\mathrm{d}a$, which reproduces the Hall response of four copies of Laughlin state at $\nu=-1/2$. At topological order level, the $U(2)_{-1,2}$ state is equivalent to $(U(1)_2\boxtimes U(1)_{-2})/\mathbb{Z}_2$, which is actually a trivial topological order, similar to the bosonic IQH state widely adopted in the construction of anyon superconductivity~\cite{Divic2025,Pichler2025}. Thus, it is natural for $z_{l,s}$ to form such a state.

Taking together, the topological response of doping $SU(3)_2$ CSL is a $U(2)_{-6,0}$ CS theory,
\beq
\mathcal{L}[A]&=&\frac{6}{4\pi}\trace{a\mathrm{d}a+\frac{2}{3}a^3}-\frac{3}{4\pi}(\mathrm{tr}\,a)\mathrm{d}(\mathrm{tr}\,a)\nn\\
&&+\frac{3}{2\pi}(\mathrm{tr}\,a)\mathrm{d}A+\frac{6}{4\pi}A\mathrm{d}A+12\Omega_g.\label{eq:SCs}
\eeq
By taking $a=\tilde{a}\mathbb{I}_2$, the self CS term of $\tilde{a}$ cancels out, suggesting Lagrangian \Eq{eq:SCs} describing a charge-$6e$ SC. Different to ones arising from doping Abelian spin liquids, this charge-$6e$ SC exhibits fermionic non-Abelian topological order described by $SO(3)_{-3}$ CS theory with chiral central charge $-9/4$~\cite{Gao2026,shi2025nonabelian}. This is because $U(2)_{-6,0}=(SU(2)_{-6}\boxtimes U(1)_0)/\mathbb{Z}_2$, and $SU(2)_{-6}/\mathbb{Z}_2=SO(3)_{-3}$. The chiral central charge of the charge-6e SC is then $c_-=6-9/4=15/4$. In addition, it hosts non-Abelian $h/(6e)$ SC vortices with quantum dimension $\sqrt{2+\sqrt{2}}$ and $\sqrt{4+2\sqrt{2}}$~\cite{Gao2026}.

\section{Discussion and Outlook}\label{sec:dis}

In this work, we have proposed and analyzed a route to charge-$6e$ superconductivity from doping $SU(3)$-symmetric spin liquids. The key organizing principle is the $\mathbb{Z}_3$ center structure of $SU(3)$ group that forbids a symmetry-preserving charge-$2e$ or charge-$4e$ condensate. The symmetry and topological properties of the resulting charge-$6e$ superconductors depend on those of the parent spin liquids.  Doping a $\mathbb{Z}_3$ quantum spin liquid can lead to a time-reversal-symmetric charge-$6e$ superconductor, which can be viewed as an $s$-wave paired state of charge-$3e$ fermionic trions. Doping Abelian chiral spin liquids gives chiral charge-$6e$ superconductors. Doping $SU(3)_1$ CSL yields a charge-$6e$ SC coexisting with residual Abelian topological order, while doping $SU(6)_1$ CSL yields a chiral charge-$6e$ SC without intrinsic topological order. Finally, doping the non-Abelian $SU(3)_2$ CSL gives a non-Abelian chiral charge-$6e$ SC intertwined with $SO(3)_{-3}$ topological order, and consequently hosts non-Abelian $h/(6e)$ superconducting vortices. Thus, within the center enforcement mechanism~\cite{Gao20264e}, charge-$6e$ SC can arise from the smaller internal symmetry group $SU(3)$, whereas the analogous route to charge-$4e$ SC typically invokes $SU(4)$ symmetry.

A natural concern is that an exact microscopic $SU(3)$ symmetry may be too restrictive for realistic materials.  We emphasize, however, that the topological ingredients used in our construction do not always require an exact microscopic $SU(3)$ symmetry.  At the level of topological quantum field theory, level--rank duality~\cite{Hsin2016,Wang2026su3} relates $SU(3)_1$, $SU(6)_1$, and $SU(3)_2$ Chern--Simons theories to alternative $U(1)_{-3}$, $U(1)_{-6}$, and $U(2)_{-3,-6}$ descriptions, respectively, up to stacking invertible fermionic state and gravitational Chern--Simons factors~\cite{WangCSL2026}. Therefore, apparently simple Abelian or non-Abelian topological states may admit dual descriptions in which the $SU(3)$ structure becomes manifest, and can serve as parent topological orders for charge-$6e$ superconductivity.  Related examples, including holon metals and superconductors constructed from dual descriptions of simple-looking topological states, have been proposed recently~\cite{Zhang2026,shi2025nonabelian}. Nevertheless, for the $SU(3)$-based mechanism to be effective in a microscopic system, some approximate three-component structure at low energy should remain important. Such a structure may arise from approximate flavor symmetry, valley or sublattice degrees of freedom, orbital degeneracy, or threefold crystalline symmetry. Recent studies on charge-$1/3$ anyon fluid obtained by doping the Jain-$3/2$ fractional Chern insulator suggests an emergent $SU(3)$ symmetry arising from the 3-flavor composite fermion~\cite{Shi2026sc,Senthil2026,Han20261,Zhang20261,Mehta2026}. From this perspective, van der Waals materials with honeycomb or triangular moir\'e structures, as well as kagome metals with strong correlation and nontrivial band topology, are natural arenas in which to search for charge-$6e$ superconductivity.

It would be valuable to establish the energetic stability of the proposed charge-$6e$ superconducting phases in microscopic models.  The bilayer $SU(3)$ Hubbard model studied here provides a natural starting point, but determining which doped phase is selected requires a quantitative treatment of microscopic interactions, effective bond and ring exchanges, and dopant dynamics. Tensor-network methods, including density-matrix renormalization group simulations, have already been applied successfully to related $SU(2)$, $Sp(4)$, and $SU(4)$ bilayer models~\cite{Oh2024,Yang2025,Oh2025,ESDreview,Gao20264e}. In the $SU(4)$ case, primary charge-$4e$ superconductivity has been verified numerically~\cite{Gao20264e}. Monte Carlo simulations of other $SU(4)$ lattice models have also revealed robust charge-$4e$ superconducting phases~\cite{Li20264e,Wan2026}. Performing analogous numerical simulations on their $SU(3)$ variants would directly test whether the proposed $\mathbb{Z}_3$ QSL and $SU(3)$-related CSL are energetically competitive, and whether their doped descendants exhibit dominant charge-$6e$ superconducting regime. Useful diagnostics include the decay of charge-$2e$, charge-$3e$, and charge-$6e$ correlation functions in quasi-one-dimensional setting, flux insertion, and superconducting stiffness~\cite{Scalapino1993,Oshikawa2000,Zaletel2014,bhler2025}. Momentum-space or entanglement signatures of trion Fermi
surfaces~\cite{Oshikawa2000,Lai2016}, and entanglement probes of the
residual Abelian or non-Abelian topological order
\cite{Kitaev2006t,Levin2006,LiHaldane2008,Zhang2012,Cincio2013,
Zaletel2013} should also be illustrating.

Experimentally, the most direct signatures of charge-$6e$ SC are flux quantization in units of $h/(6e)$, Josephson oscillations with frequency $6eV/h$, and Little--Parks oscillations with period $h/(6e)$.  A charge-$6e$ SC should also be distinguished from an ordinary charge-$2e$ SC by the absence of long-range charge-$2e$ and charge-$4e$ order parameters, while charge-$6e$ hextet remains coherent. For chiral charge-$6e$ SC, thermal Hall response and chiral edge modes can further diagnose the underlying topological order. In the non-Abelian case, the $h/(6e)$ vortices themselves are non-Abelian, whose fusion degeneracies and braiding properties provide a sharp distinction from both conventional superconductors and Abelian higher-charge superconductors, and may offer a route toward topological quantum information processing~\cite{Shi2026}.

\section{Acknowledgments}

We acknowledge Clemens Kuhlenkamp, Zhaoyu Han, Taige Wang, Zhehao Dai, Ziqiang Wang, Zhengyan Darius Shi, T. Senthil, Alexander Seidel, and especially Ya-Hui Zhang for helpful discussions. YQW is supported by the JQI postdoctoral fellowship at the University of Maryland. BZ is supported by a startup fund from the Johns Hopkins University. ZQG is supported through the Berkeley graduate program. 

{\it Note added.} Upon the completion of this work, we became aware of another study on charge-$6e$ SC based on $SU(3)$ group~\cite{6e2}. While both works adopt similar symmetry perspective, the microscopic models and setups are different.

\bibliography{6e.bib}

\appendix

\section{Transformation of gauge fields}\label{app:A}

Starting from Lagrangian \Eq{eq:Lraw}, the $GL(7,\mathbb{Z})$ transformation $(\tilde{\alpha},{\bm \alpha})=(a,{\bm \beta})W^\mathbf{T}$ is
\beq
W=\begin{pmatrix}
2 & 0 & 0 & 0 & 0 & 0 & -1 \\
-1 & 0 & 0 & 0 & 0 & 0 & 0 \\
-1 & 1 & 0 & 0 & 0 & 0 & 0 \\
-1 & 0 & 1 & 0 & 0 & 0 & 0 \\
-1 & 0 & 0 & 1 & 0 & 0 & 0 \\
-1 & 0 & 0 & 0 & 1 & 0 & 0 \\
-1 & 0 & 0 & 0 & 0 & 1 & 0
\end{pmatrix}.
\eeq
After transformation, the resulting $K$-matrix is
\beq
K_{\mathrm{res}}=\begin{pmatrix}
-2 & 0 & 0 & 0 & 0 & 1 \\
0 & -2 & 0 & 0 & 0 & 1 \\
0 & 0 & -2 & 0 & 0 & 1 \\
0 & 0 & 0 & -2 & 0 & 1 \\
0 & 0 & 0 & 0 & -2 & 1 \\
1 & 1 & 1 & 1 & 1 & -3
\end{pmatrix},
\eeq
which has central charge $c_-=-6$ and torus ground state degeneracy $16$. We adopt the sign convention for $K$-matrix and $SU(N)_k$ CS term as
\beq
\mathcal{L}=-\frac{K_{IJ}}{4\pi}\alpha_I\mathrm{d}\alpha_J-\frac{k}{4\pi}\trace{a\mathrm{d}a+\frac{2}{3}a^3}.
\eeq

\section{Parton mean-field phase diagram at doping $x=1$}\label{app:B}

In this section we map out the mean-field phase diagram of the effective $SU(3)$ spin model at $x=1$ using parton construction \Eq{eq:parB}. The effective Hamiltonian is the $J$--$K_3$ model~\cite{Lai2013}, with $J\sim t^2/u$ and $K_3\sim t^3/u^2$ in \Eq{eq:bH},
\beq
   H_{x=1}&=&J\sum_{\langle \mathbf{ij}\rangle} P_\mathbf{ij}
   + K_{3}\sum_{\mathbf{ijk}\in\triangle}\Bigl(e^{i\varphi}\,P_\mathbf{ijk} + \mathrm{h.c.}\Bigr)\nn\\
   &&+ K_{3}\sum_{\mathbf{ijk}\in\bigtriangledown}\Bigl(e^{-i\varphi}\,P_\mathbf{ijk} + \mathrm{h.c.}\Bigr),
\label{eq:HJK}
\eeq
where $P_\mathbf{ij}$ and $P_\mathbf{ijk}$ are the bond and plaquette exchange terms,
\begin{equation}
\begin{split}
   P_\mathbf{ij}=&\sum_{\alpha,\beta}
   d^{\dagger}_{\alpha}(\mathbf{i})d_{\beta}(\mathbf{i})\,
   d^{\dagger}_{\beta}(\mathbf{j})d_{\alpha}(\mathbf{j}),\\
   P_\mathbf{ijk}=&\sum_{\alpha,\beta,\gamma}
   d^{\dagger}_{\alpha}(\mathbf{i})d_{\beta}(\mathbf{i})\,
   d^{\dagger}_{\beta}(\mathbf{j})d_{\gamma}(\mathbf{j})\,
   d^{\dagger}_{\gamma}(\mathbf{k})d_{\alpha}(\mathbf{k}),
\end{split}
\label{eq:permops}
\end{equation}
with $(\mathbf{i},\mathbf{j},\mathbf{k})$ ordered counterclockwise on up ($\triangle$) and down ($\bigtriangledown$) triangles, subject to the local constraint $\langle n_{d}(\mathbf{i})\rangle=1$. Note that the ring phase $\varphi$ in \Eq{eq:HJK} derived from \Eq{eq:bH} in the main text should be $\varphi=0$ to preserve time-reversal symmetry. Nevertheless, in the mean-field calculation we take it as a tunable coefficient, and view nonzero $\varphi$ state as competing states of CSL states arising at $\varphi=0$.

We consider the mean-field Hamiltonian,
\beq
H_{\rm MF}=-\sum_{\langle \mathbf{ij}\rangle}\sum_{\sigma}
\Bigl(\chi_\mathbf{ij}\,d^{\dagger}_{\sigma}(\mathbf{i})d^{\phantom{\dagger}}_{\sigma}(\mathbf{j})+ {\rm h.c.}\Bigr),\label{eq:HMF}
\eeq
labeled by the gauge-invariant flux pair, $(\Phi_{\triangle},\Phi_{\bigtriangledown})$,
\begin{equation}
  \Phi_{\triangle(\bigtriangledown)} =\mathrm{arg}\prod_{\langle\mathbf{ij}\rangle\in\triangle(\bigtriangledown)}^{\circlearrowleft}\chi_\mathbf{ij}.
  \label{eq:fluxdef}
\end{equation}
Here the sum of the hopping phases taken counterclockwise around each elementary
triangle. Specifically, we compare the uniform-flux states 
$\varphi=\Phi_{\triangle}=\Phi_{\bigtriangledown}=n\pi/3$ with $n=0,\pm1,\pm2,3$, together with
the trimer state---a product of three-site $SU(3)$ singlets on a triangle
covering, which breaks translational symmetry and corresponds to the
decoupled-triangle ansatz whose hopping amplitude is nonzero only on the
bonds of the chosen covering. Owing to the $\mathcal{T}\circ C_{2}$ symmetry
noted above, ansatzes with opposite fluxes $\pm\Phi$ are exactly degenerate
at every $(K_3,\varphi)$; the phases below are therefore labeled by
$|\Phi_{\triangle}|$, with the chirality selected spontaneously.

For each ansatz we evaluate the Gibbs--Bogoliubov variational free energy
$F=\langle H\rangle_{\rm MF}+F_{\rm MF}-\langle H_{\rm MF}\rangle$, where
$\langle H\rangle_{\rm MF}$ is obtained by Wick's theorem from the thermal
bond and density expectation values of \Eq{eq:HMF}, with the chemical
potential fixed by the filling constraint; at $T=0$, $F$ reduces to the variational energy. The Wick functional
of the two representations differs only in the sign of the density-type contractions. 

\begin{figure}[htbp]
    \centering
    \includegraphics[width=\linewidth]{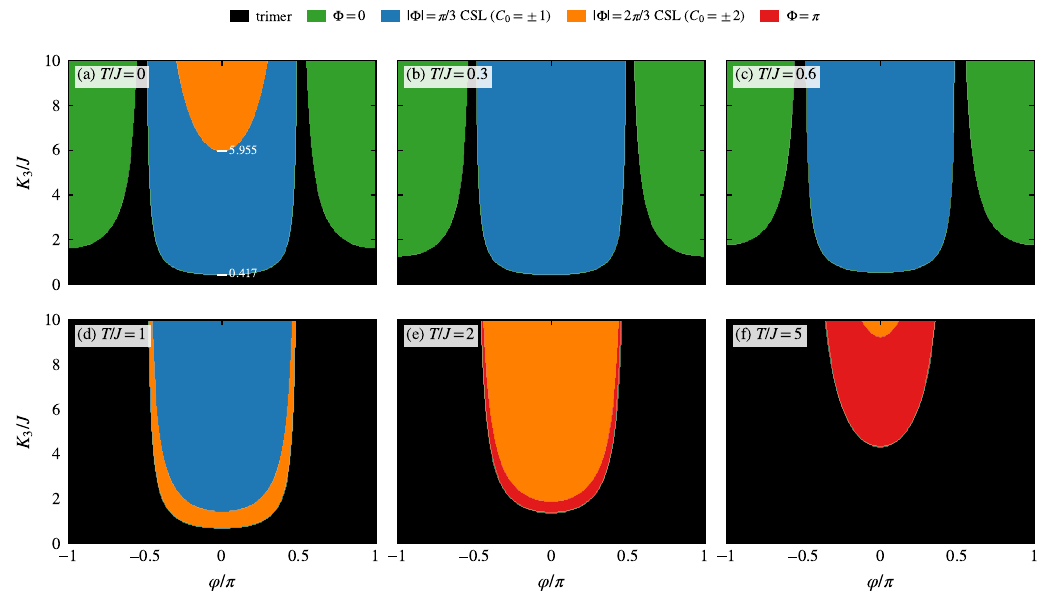}
    \caption{Mean-field phase diagram of \Eq{eq:HJK} in the $(\varphi,K_3)$ plane at temperatures $T/J=0$--$5$ [panels (a)--(f)].
    Colors label the lowest energy ansatz: trimer (black), $\Phi=0$ (green),
    $|\Phi_{\triangle}|=\pi/3$ CSL with $C=\pm1$ (blue),
    $|\Phi_{\triangle}|=2\pi/3$ CSL with $C=\pm2$ (orange),
    and $\Phi=\pi$ (red). White markers in (a) denote the $\varphi=0$
    boundaries $K_3/J=0.417$ and $5.955$ of Ref.~\cite{Lai2013}. Calculations use a
    $24\times24$ unit-cell lattice.}
    \label{fig:fermionPD}
\end{figure}

The resulting phase diagrams are shown in Fig.~\ref{fig:fermionPD} as a function of $\varphi$, $K_3/J$, and $T/J$. Here $T$ is temperature, where finite temperature phase diagrams suggest competing phases of CSL. For $\varphi=0$ and $T=0$, which is mostly concerned in this work, the ground state evolves with increasing $K_3$ from the trimer state (valence bond solid or VBS) to the $\Phi_{\pi/3}$ CSL at $K_3/J\simeq0.42$ with Chern number $C=\pm 1$, and then to the $\Phi_{2\pi/3}$ CSL at $K_3/J\simeq5.96$ with $C=\pm 2$, consistent with Ref.~\cite{Lai2013}. For finite temperature, a $\Phi=\pi$ state emerges as a competing phase of CSL. Such a $\pi$-flux state realizes a $U(1)$ spin liquid. Upon doping, fermi surface of $d_\sigma$ parton is formed. Detailed study of doping the VBS and the $U(1)$ spin liquid is left for future work. At finite $\varphi$ the chiral-spin-liquid dome extends to $|\varphi|\lesssim 0.45\pi$, beyond which the trimer state and the $\Phi=0$ ansatz are stablized. Near $\varphi=\pm\pi/2$, where the ring-exchange
energy gain vanishes, the trimer state wins at all $K_3$. We also perform parton mean-field calculation using parton construction \Eq{eq:parA}, where we find $\left<h_\sigma(\mathbf{i})\right>\neq 0$ at $x=1$ and $T=0$.


\end{document}